\documentclass[twoside,leqno,twocolumn]{article}

\usepackage{ltexpprt}
\usepackage{hyperref}
\usepackage{microtype}

\usepackage[utf8]{inputenc}
\usepackage[english]{babel}
\usepackage[backend=bibtex,maxcitenames=1]{biblatex}
\usepackage{graphicx}
\graphicspath{ {./images/} }

\usepackage{algpseudocode}
\usepackage{algorithm}

\usepackage{caption}
\usepackage{enumitem}

\begin{document}

\title{\Large Guaranteed, Predictable, Polynomial AGV Time-Pathing}
\author{James Forster}

\date{}

\maketitle

\begin{abstract} \small\baselineskip=9pt

In this paper we present a framework of key algorithms and
data-structures for efficiently generating timetables for any number
of AGVs from any given positioning on any given graph to accomplish
any given demands as long as a few easily satisfiable assumptions are
met. Our proposed algorithms provide guaranteed solutions in
predictable polynomial running-times, which is fundamental to any
real-time application. We also develop an improved geographic
reservation algorithm that provides a substantial run-time improvement
of the previously best-known algorithm from $O(nm)$ to $O(n)$.

\end{abstract}

\section{Introduction}
\label{sec:introduction}

The use of Automated Guided Vehicles is widespread in modern industry,
and the effective operation of AGVs is a complex problem with effects
on the productivity of the system as a whole. In this paper we focus
on the centralised generation of timetables for AGVs travelling on a
graph in order to complete a set of online transportation requests
called demands using time-pathing methods.

No existing methods that attempt to address this problem have
predictable polynomial run-time due to having non-zero failure rates
which mean they may have to be ran an unbounded number of times before
a solution is found. This makes such approaches unsuitable for
real-time applications in enviroments such as automated warehouses and
automated shipping yards where demands may be updated in real-time.
Hence, a solution must be provided in a reasonable and predictable
timeframe in order for the system to maintain high availability
without manual intervention. In this paper we propose such a method
that will always succeed given a few assumptions are met.

In section \ref{sec:preliminaries} we discuss the definitions used
throughout the paper, the modelling of the problem, the key
assumptions that our later algorithms rely upon and related work.
In section \ref{sec:gap_query_tree} we describe a new data-structure
that is optimised for the key operations used during time-pathing.
In section \ref{sec:guaranteed_time_pathing} we propose a new algorithm
for time-pathing AGVs that guarantees a viable timetable in
predictable running-time.
In section \ref{sec:optimisations_for_large_graphs} we show how two
classical graph search optimisation techniques, guided search and
partial search, can be applied to the time-pathing algorithm.
In section \ref{sec:geographic_reservations} we describe how edge
subdivision can be used to increase resource utilisation efficiency
and propose an algorithm that improves the running-time of geographic
reservations \cite{stenzel4,stenzel8} from $O(nm)$ to $O(n)$ by
exploiting the symmetric nature of geographic links.
Finally, in section \ref{sec:simulations} we show the results of
simulations which demonstrate the performance of our proposed
algorithms on a basic grid graph which show their initial performance
characteristics.

\section{Preliminaries}
\label{sec:preliminaries}

\subsection{Definitions}
\label{sec:definitions}

\begin{itemize}[noitemsep,nolistsep]
	\item{AGV: Automated Guided Vehicle}
	\item{Demand: A request to move an item from a given pickup node to a given drop-off node}
	\item{Assignment: The pairing of an AGV to a demand}
	\item{Resources: Nodes + Edges}
	\item{Reservation: An interval of time associated with an AGV and the resource they made it on}
	\item{Time-Window: An interval of time available to an AGV on a given resource by not being reserved by any other AGV}
	\item{Occupation-Interval: An interval of time spent directly on a resource by an AGV}
	\item{Geographic Reservation: A reservation made simultaneously on geographically nearby resources to a base resource in order to maintain AGV safety buffers while time-pathing}
	\item{Anchor Node: A node on the graph designated for allowing AGVs to park on them indefinitely, also known as parking nodes}
	\item{Time-Graph: A graph with AGVs' reservations for each resource in the graph}
	\item{Time-Path: A path in the graph with associated occupation-intervals for each visited resource}
	\item{Time-Pathing: A method for finding the fastest time-path in a time-graph}
	\item{TimeTable: The time-paths associated with each AGV that specify where on the graph each AGV will be at any given time}
	\item{Edge Subdivision: An operation on a graph where you split each edge into a number of smaller edges by inserting new nodes}
	\item{Safety Buffer: A minimum separation distance maintained between the centres of all AGVs to prevent physical collisions}
\end{itemize}

\subsection{Modelling and Key Assumptions}
\label{sec:modelling_and_key_assumptions}

The model we assume throughout this paper is that demands are given in
a sequence where each demand has an associated horizon time when it
becomes known. This online model is a generalisation of the offline
model where all demands have a horizon time of zero. Online demands
are common where demands have a dependence on real-time conditions
such as customer orders.

We assume AGVs are capable of following their time-paths exactly and
that they never break down as well as no other changes to the graph
such as blockages, however, the framework proposed is easily
extensible to manage such faults, see
\cite{maza1,maza5,mors10,stenzel8}. AGVs are assumed to all be of the
same size and move at the same speed and acceleration is not modelled,
however such extensions are possible, see
\cite{stenzel4,stenzel8,lienert18b}.

We also assume that we are given a mixed graph $G = (N, E)$, where $N$
is the set of nodes and $E$ is the set of edges, that represents the
layout that the AGVs can move on. This is common in environments such
as warehouses where there is a fixed layout of aisles or where there
are humans working alongside the AGVs and so predictable movement
paths are required for safety reasons.

We impose the following restrictions onto $G$ as the basis for
allowing the proposed algorithms to work:

\begin{enumerate}[noitemsep]
	\item\label{itm:strongly_connected} $G$ is strongly connected
	\item\label{itm:anchor_node_magnitude} There exists a set of anchor nodes $A \subseteq N$ such that there are at least as many anchor nodes as AGVs
	\item\label{itm:subgraph_strongly_connected} The subgraph obtained from $G$ by removing all nodes in $A$ is also strongly connected \cite{drotos}
	\item\label{itm:no_edges_between_anchors} There exist no edges in $E$ between two nodes both in $A$
	\item\label{itm:demands_not_on_anchors} No demands have pickup or drop-off locations in $A$
\end{enumerate}

\ref{itm:strongly_connected}, \ref{itm:subgraph_strongly_connected}
and \ref{itm:demands_not_on_anchors} are required so that no matter
where on the graph an AGV is positioned it will be able to carry out
any demand without being blocked by parked AGVs.
\ref{itm:anchor_node_magnitude} and \ref{itm:no_edges_between_anchors}
are requirements for the algorithms in sections
\ref{sec:anchored_time_pathing} and \ref{sec:anchorisation} respectively.

The problem is then to produce a timetable for coordinating multiple
AGVs such that they complete their assigned demands while avoiding
collisions by maintaining a given safety buffer.

\subsection{Related Work}
\label{sec:related_work}

The concept of conflict-free AGV routing was first proposed by
\cite{broadbent}, they kept a global timetable of node-occupation
times and then used Dijkstras to find the shortest spatial path for an
AGV. This was then converted into a time-path and the new
node-occupation times were calculated. If these new node-occupation
times conflicted with any existing node-occupation times in the
timetable then a set of heuristic rules were used to generate a new
time-path depending on the type of collision detected: head-on,
catch-up or crossing collisions.

\cite{huangs1,huangs2,tanchocos} proposed labeling algorithms which
guarantee the shortest time-path through any given time-graph if one
exists. \cite{huangs1,huangs2} assume all resources of a graph can
only be occupied by one AGV at any one time whereas \cite{tanchocos}
considers only nodes to be exlusive and allows multiple AGVs on edges
as long as they do not conflict with one another. This does mean
\cite{tanchocos} may produce more optimal time-paths than
\cite{huangs1,huangs2}, but it comes at the cost of a higher
computational complexity, $O(n^2\log{n})$ and $O(n^2v^2)$ respectively
where $n$ is the number of time-windows in the time-graph and $v$ is
the number of AGVs. Both algorithms, however, rely upon temporal
separation between AGVs which only guarantees spatial separation when
every edge in the graph is sufficiently sized compared to the size of
the AGVs.

\cite{stenzel4} extended \cite{huangs1,huangs2,tanchocos} to add
geographic reservations that ensure spatial separation by reserving a
set of geographically linked resources at any given time rather than
just one. \cite{stenzel8} introduced the idea of initialising the
labelling algorithm with multiple starting labels which has the effect
of selecting the closest source label to the destination with the same
worst-case run-time complexity as the single-source version.

Since AGV demands involve picking up and then dropping off items it is
necessary to visit multiple sequential destinations. \cite{mors10}
also extended \cite{huangs1,huangs2,tanchocos} to provide the shortest
time-path for a sequence of destinations, since unlike spatial
shortest path-finding simply concatenating the individual shortest
time-paths between destinations together does not always work.
\cite{brownlee} later reproduced an equivalent algorithm.

\cite{mors10,zutt} proposed a slight variation of the algorithm
proposed by \cite{huangs1,huangs2} which has a much improved
worst-case complexity since it proves that the first time you
reach-out to a new time-window it is guaranteed to be the fastest you
can reach that time-window. This means that unlike
\cite{huangs1,huangs2} not only do you only reach-out from each
time-window once but each time-window is also only reached-out to
once. It runs in $O((m + n)v\log{n})$ where $m$ is the number of
connections in the graph between resources. A major caveat to this
algorithm is that unlike \cite{huangs1,huangs2} it is only complete if
no heuristic guidance function is used. This can mean in practise that
it is slower than \cite{huangs1,huangs2}'s algorithm when a heuristic
is used despite a lower worst-case time-complexity.

\cite{stenzel8,mors10} suggest a basic framework for using an
underlying time-pathing algorithms to produce timetables which fulfill
given demands: AGVs are iteratively time-pathed from their current
location to the pickup location of their next assigned demand in the
global time-graph, then they are similarly time-pathed to the drop-off
location of that demand. Finally, the occupation-intervals produced
from each time-path are reserved back onto the global time-graph to
prevent collisions. This process then repeats until all demands have
been time-pathed and a timetable has been created.

Assuming a given assignment of AGVs to demands, the order in which you
time-path AGVs can have a large effect on the quality of the produced
timetable. \cite{mors11,rybecky18,rybecky20} explore the effects of
various methods for selecting the order in which to time-path AGVs to
optimise various timetable metrics. Heuristics such as longest-first
and simply random-order were particularly effective depending on the metric
measured.

Unfortunately, \cite{stenzel8,mors10,mors11,rybecky18,rybecky20} all
suffer from the same effect that the time-pathing of multiple AGVs to
complete their assigned demands can fail when there exists no possible
time-path to one of their destinations. In \cite{stenzel8} it is stated
that only permanent reservations lead to failures in time-pathing,
which is not correct as temporary reservations can also cause failures
if they isolate an AGV's source time-window.
\cite{rybecky18,rybecky20} incorporates a success rate as part of
their analysis of time-path ordering as a measure of quality, which
suggests a great lack of guarantees in realising any given set of
demands in a reasonable amount of time. In \cite{mors10,mors11} always
available, regardless of the number of reservations on them, source
and destination resources are introduced to prevent failures but such
assumptions are unrealistic in practise.

\cite{sem} comes much closer to achieving guaranteed failure-free
time-pathing. They use the concept of parking nodes, special nodes
designed specifically for AGVs to park and hence reserve indefinitely
while they are idle. Then after completing their set demand AGVs are
always sent to an available non-reserved parking node so as to not get
in the way of other AGVs and such cause a failure. Unfortunately,
\cite{sem} still reports failures in time-pathing where they
theoretically should not ever fail and it is put down to an bug in the
implementation, but we believe it is due to the lack of multiple
sequential destination time-pathing with their parking node system.

The idea of using parking nodes to start and end time-paths was also
proposed in \cite{drotos,malopolski} however they do not use
time-pathing methods and instead use simple spatial path-finding to
find paths for AGVs to take and then use a resource claiming algorithm
to provide deadlock and conflict-free execution of the found paths by
AGVs. This results in each AGV taking the shortest spatial distance
paths that often take longer to travel than longer spatial paths due
to needing to wait for other AGVs in busy areas.

There is also the issue of what to do if you are starting from a
positioning of AGVs where some of the AGVs are not on parking nodes so
attempting to start time-pathing immediately could fail. \cite{csutak}
suggest a method of reserving the initial resources where all AGVs are
located for an amount of time $x$ and attempting to time-path each AGV
where each AGV considers its own initial reservation as a time-window
but respects other AGVs' initial reservations. Then the value of $x$
is increased every time the initialisation fails until a solution is
found. This brute force method is not suitable for applications
requiring predictable performance as its execution time is unbounded.

\section{Gap-Query Optimised Interval-Tree Data-Structure}
\label{sec:gap_query_tree}

A key part of the underlying time-pathing algorithm is its use of
time-intervals to signify when resources are either reserved by an AGV
or free. To do this efficiently we store AGVs' reservations and use
the gaps between these reservations to define the time-windows. This
data-structure needs to support three key operations: insertion of new
reservations, removal of existing reservations and gap-queries to find
time-windows in a given range. Insertions are done after a new viable
time-path has been found to prevent collisions. Removals are sometimes
necessary when cancelling a previously inserted time-path. Finally,
gap-queries are used during time-pathing in order to find new viable
time-paths in a time-graph.

Additionally, we require insertions, removals and gap-queries to take
an identifier for the AGV the intervals should be associated with,
this is used in gap-queries so that AGVs see their own reservations as
gaps as an AGV cannot collide with itself and hence it can violate its
own reservations. This property is a requirement for the algorithms in
section \ref{sec:anchorisation}.

Importantly, multiple reservation-intervals associated with the same
AGV will never overlap one-another since an AGV cannot reserve the
same time more than once as that would not make any sense. However,
different AGVs' reservation-intervals can overlap one-another. This
property is a requirement for section
\ref{sec:geographic_resource_linking}.

Following from these requirements we propose a data-structure that
uses the fact no interval for the same AGV overlaps itself in order to
store all AGVs' intervals in a single self-balancing non-overlapping
tree data-structure such as a red-black tree.

\subsection{Insertions}
\label{sec:insertions}

To insert a new reservation into the data-structure we add the AGV's
identifier to the sets associated with the intervals that overlap the
new reservation-interval. We also insert a new interval for any
non-covered sections of the new reservation-interval with a new
associated set containing just the new AGV's identifier. Finally, care
must be taken to split intervals that partially overlap the end-points
of the new reservation-interval so as to only change the overlapping
portions of the interval-tree.

\begin{figure}[h!]
\centering
\includegraphics[width=7cm]{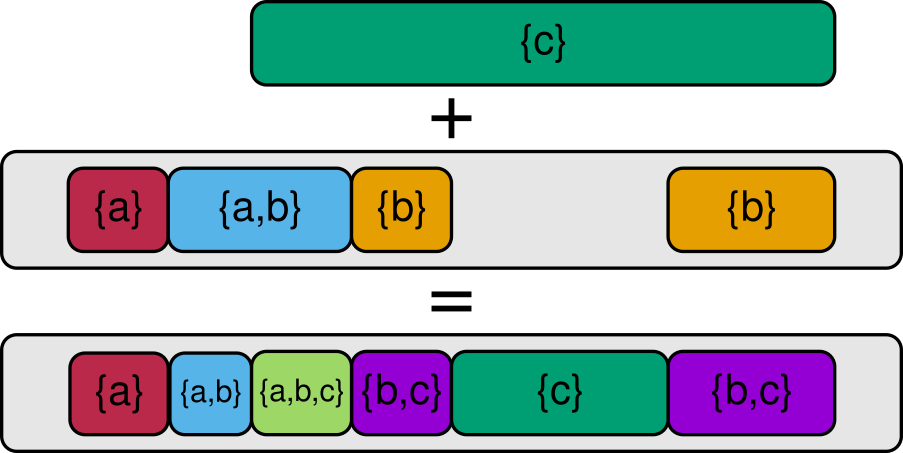}
\captionsetup{justification=centering}
\caption{An example of an insertion into an interval-tree}
\end{figure}

Finding the first overlapping interval takes $O(\log{n})$ where $n$ is
the number of intervals in the tree. Assuming that hash-sets are used
for storing AGV identifiers then they can be updated in $O(1)$.
Iterating in ascending order across the tree for every interval that
overlaps the removal reservation-interval updating each interval or
inserting new intervals can be done in $O(k)$ where $k$ is the number
of intervals that overlaps the inserted interval.

Therefore this has a run-time complexity of $O(\log{n} + k)$.

\subsection{Removals}
\label{sec:removals}

Removals do the inverse of insertions by removing the associated AGV
identifier from all intervals in the tree that overlap the given
removal reservation-interval splitting intervals as appropriate. Care
must also be taken to re-merge intervals that touch and have the same
set of associated AGV-intervals to prevent fragmentation.

\begin{figure}[h!]
\centering
\includegraphics[width=7cm]{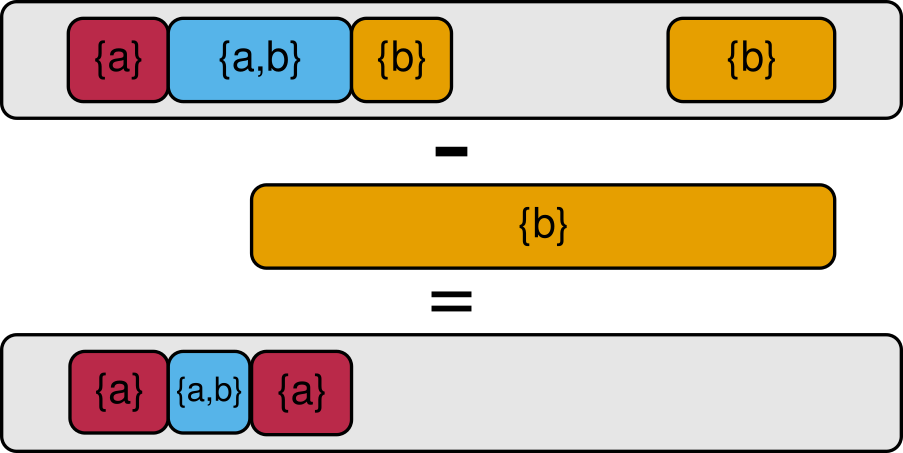}
\captionsetup{justification=centering}
\caption{An example of a removal from an interval-tree}
\end{figure}

Removals can be implemented in a similar fashion to insertions with
first an $O(\log{n})$ lookup followed by $O(k)$ interval updates and so has
an identical run-time complexity of $O(\log{n} + k)$.

\subsection{Gap-Queries}
\label{sec:gap_queries}

To find all gaps for in a given interval with a given AGV identifier
we simply examine all overlapping intervals in the tree and filter out
those that contain AGV specifiers not equal to the given AGV
specifier. We also consider gaps in the stored intervals and
merge touching gap-intervals and viable same-AGV intervals.

\begin{figure}[h!]
\centering
\includegraphics[width=7cm]{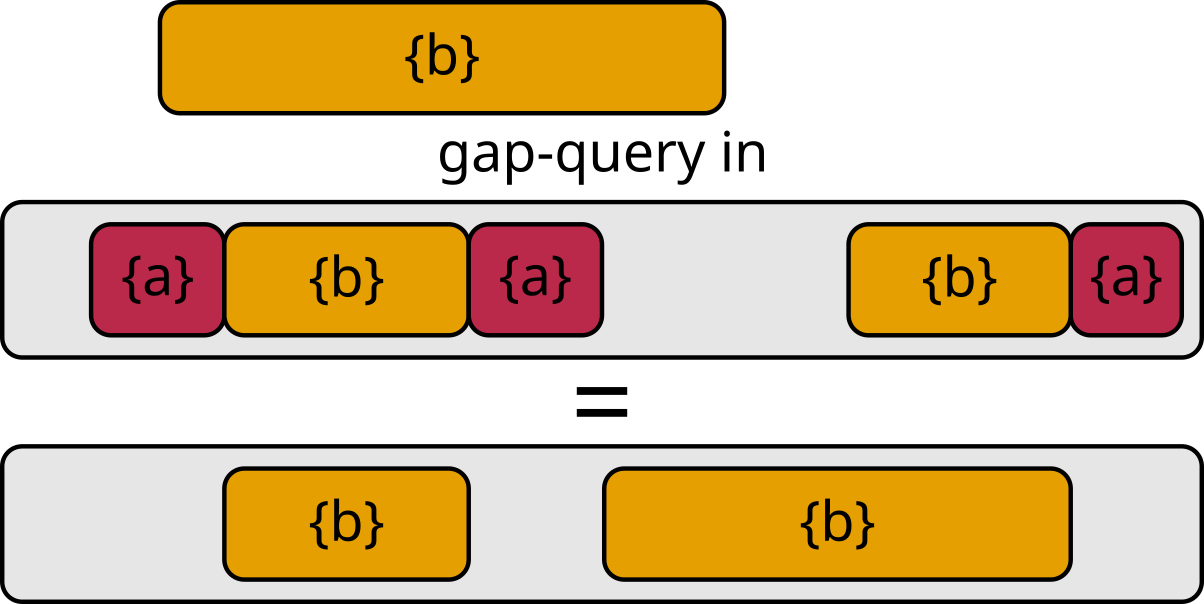}
\captionsetup{justification=centering}
\caption{An example of a gap-query on an interval-tree}
\end{figure}

In the same fashion as insertion and removal this can be implemented
by first a $\log{n}$ lookup followed by $k$ iterations across
overlapping intervals and gaps merging them where appropriate and
hence it too has a run-time complexity of $O(\log{n} + k)$.

\subsection{Conclusion}
\label{sec:interval_conclusion}

This data-structure has been implemented in the Rust programming
language and published with an open-source license on GitHub,
available at \url{https://github.com/ripytide/gap_query_interval_tree}.

With a data-structure for storing the reservations of AGVs' it is then
trivial to represent a time-graph by associating each resource with its
own reservation interval-tree.

\section{Guaranteed Time-Pathing}
\label{sec:guaranteed_time_pathing}

\subsection{Additional Time-Pathing Features}
\label{sec:additional_time_pathing_features}

The following algorithms all make use of a fundamental sub-procedure
called time-pathing, initially proposed by
\cite{huangs1,huangs2,tanchocos}, but they also require a number of
additional features:

\begin{enumerate}[noitemsep]
	\item\label{minimum_stop_duration} Minimum stop duration at the destination \cite{lienert17}
	\item\label{multiple_source_labels} Multiple source labels \cite{stenzel8}
	\item\label{multiple_sequential_destinations} Multiple sequential destination nodes \cite{mors10}
	\item\label{multiple_parallel_destinations} Multiple parallel destination nodes
	\item\label{starting_from_an_edge} Starting from any position along an edge
\end{enumerate}

Our implementation of the time-pathing algorithm allows all of these
features in any combination.

\subsection{Anchored Time-Pathing}
\label{sec:anchored_time_pathing}

As described in \cite{drotos,malopolski,sem}, time-pathing should
always ends at an anchor node, we call this anchored time-pathing.
Similarly, we call a timetable in which every AGV ends at an anchor
node an anchored timetable. If the timetable is always anchored then
any AGV will always be able to fulfill any demands.

An easy way to prove that this is the case is for an AGV to simply
wait at its anchor node until all other AGVs have also entered their
anchor nodes for the final time. At which point the AGV can then
travel freely on the subgraph of the base graph minus the anchor nodes
to fulfill its required demands, see assumption
\ref{itm:subgraph_strongly_connected} in section
\ref{sec:modelling_and_key_assumptions}, before returning to an anchor
node to maintain the anchored status of the timetable.

However, this is not a very efficient use of time as often there will
exist a much faster time-path for an AGV to complete its assigned
demands before all other AGVs reach their anchor nodes. This is
precisely what the time-pathing algorithm stated in section
\ref{sec:additional_time_pathing_features} will find.

Anchored time-pathing by itself does not however solve the issue of how we
obtain a so-called anchored timetable to begin with.

\subsection{Anchorisation}
\label{sec:anchorisation}

In this section we propose two algorithms to initialise an anchored
timetable from any given AGV positioning on a graph such that anchored
time-pathing may begin. We assume no two AGVs are initially positioned
on the same resource.

The first step of both algorithms is to initialise the time-graph such
that the resources that AGVs are positioned on are reserved for
themselves indefinitely. This prevents other AGVs from colliding with
them while they are initially stationary.

The next step is to time-path each AGV from its current resource to
any anchor node. When time-pathing we use feature
\ref{minimum_stop_duration} to require a minimum stop duration of
infinity since to properly anchor an AGV it must have an indefinite
reservation and not just a finite one.

We propose two different methods for doing this, so-called naive and
greedy anchorisation.

The naive method simply loops through each AGV and attempts to
time-path it to any anchor node using feature
\ref{multiple_parallel_destinations} of time-pathing.

Since in the worst case only one AGV may ancho at a time, this has a
worst-case run-time complexity of $O(n^2t)$, where $n$ is the number
of AGVs and $t$ is the run-time complexity of the used time-pathing
algorithm, since in the worst case you would have to attempt to anchor
every AGV once and then again after successfully anchoring any AGV.

The greedy method uses features \ref{multiple_source_labels} and
\ref{multiple_parallel_destinations} to time-path from every AGVs
source location to every anchor node simultaneously. Since this
produces the fastest time-path for any AGV to any available anchor
node it will always produce a viable time-path to anchor an AGV. You
can simply repeat this process as many times as you have AGVs until
every AGV has been anchored.

This has a worst-case run-time complexity of $O(nt)$ since you only
need to time-path $O(n)$ times and as shown in \cite{stenzel8} using
feature \ref{multiple_source_labels} of time-pathing has no effect on
the worst-case run-time complexity.

\section{Optimisations for large graphs}
\label{sec:optimisations_for_large_graphs}

\subsection{Guided Search}
\label{sec:guided_search}

The run-time for regular time-pathing can scale quite poorly with the
size of the graph due to it searching in all directions at once
indiscriminately similar to Dijkstras. By using a heuristic function
as suggested in \cite{huangs1,huangs2,tanchocos,mors10} we can gain a
significant speed improvement in nearly all scenarios. This is
equivalent to how A* is Dijkstras but using heuristic function.

Unlike A*, we must also account for multiple sequential destination
time-pathing, feature \ref{multiple_sequential_destinations}, in our
heuristic functions where given the current label and the list of
remaining destinations they must give an expected remaining cost for
the remainder of the whole route from the current label.

\subsection{Partial Search}
\label{sec:partial_search}

Following from the original brute-force strategy suggested in
\cite{broadbent} of doing a pre-pass of spatial pathing and then
checking if the found spatial path contains a viable conflict-free
time-path in the time-graph, \cite{smolic} proposed an algorithm that
could produce any viable time-path on a given spatial path via a
process of checking and adjusting time-windows. Unfortunately, no
time-complexity was given, instead however, we propose simply re-using
the standard time-pathing algorithm but only inputting the sub-graph
obtained by only using resources from the selected spatial path,
which will produce an equivalent result.

We also propose a method for generating this minimal sub-graph such
that time-pathing will always succeed when using anchored time-pathing
and the multiple sequential destination feature
\ref{multiple_sequential_destinations}.

We start this process by initialising an new empty sub-graph in which
we will add a number of resources from the full base graph in order to
ensure a viable time-path exists in the sub-graph.

Next, following the base principal of anchored time-pathing we must
ensure that the AGV retains its anchoring time-path by adding all
resources used in the previously planned anchoring time-path.

Finally, we must ensure that there exists a time-path from the AGVs
anchor node to every destination in its multiple sequential
destination route. We suggest two strategies one could use here, the
first is a star-topology by adding the shortest spatial path from the
anchor node to each destination node individually, forming a star. The
second strategy is a chain-topology by adding the shortest spatial
path from the anchor node to the first destination node in the route
and then the shortest spatial paths between each sequential pair of
destination nodes in the route, forming a chain. When finding these
spatial paths it is crucial that they cannot include other anchor
nodes since all anchor nodes may be indefinitely unavailable if other
AGVs are parked on them.

The partial graph topology strategy used is likely to have a large
effect on the quality of solutions, for example, a star-topology is
likely to force an AGV to travel back to its anchor node between every
destination whereas a chain-topology may allow an AGV to take a more
direct and therefore faster path. Also, the performance of the
spatial pathing algorithm used may have an impact on the
overall runtime since spatial paths must generated multiple
times per each time-pathing to generate the input sub-graph.

\section{Geographic Reservations}
\label{sec:geographic_reservations}

\subsection{Geographic Resource Linking}
\label{sec:geographic_resource_linking}

The time-pathing used so far only provides the guarantee that no AGVs
will attempt to occupy the same resource at the time via resource
reservations in the time-graph, however, this does not imply physical
separation as for example two AGVs can be arbitrarily close to one
another by occupying different edges on either side of the same node.

This issue is solved by using geographic reservations
\cite{stenzel4,stenzel8}, which allow us maintain physical separation
between AGVs by linking geographically nearby resources, such as the
two adjacent edges from the previous example, so that when the centre
of an AGV occupies a resource they also reserve those geographically
nearby resources which prevents other AGVs from time-pathing too
close to one another.

\subsection{Edge Subdivision}
\label{sec:edge_subdivision}

When the length of edges in a graph are large in proportion to the
physical size of AGVs it can lead to inefficient resource utilisation
where AGVs are forced to reserve entire large edges when they may only
need to reserve small portions of them to uphold the safety buffer.

To solve this, a pre-processing step can be applied to the given input
graph before any anchorisation or time-pathing is performed in order
to subdivide all large edges until they are suitably small in
comparison to AGVs. There are likely diminishing rewards to the amount
of edge subdivisions used on solution quality as resource utilisation
efficiently approaches 100\%.

Importantly, the key assumptions are upheld when subdividing the edges
of any valid input graph meaning no changes to the anchored
time-pathing completeness guarantees for the subdivided graph.

\subsection{Generating Geographic Reservations}
\label{sec:generating_geographic_reservations}

The algorithm presented for generating the sets of geographic
reservations for a given time-path in \cite{stenzel8}, shown in
Algorithm \ref{alg:naive}, runs in $O(nm)$ where $n$ is the number of
time-windows in the time-path and $m$ is the average number of
geographically linked resources that each resource has. We propose an
equivalent algorithm that cuts this down to $O(n)$ by exploiting the
strongly connected and symmetric structure of geographically linked
resources, shown in Algorithm \ref{alg:boundary}.

In the naive algorithm every interval in the base time-path must be
duplicated for every linked geographic resource, but since often when
an AGV moves onto the next resource in its time-path that resource
will still be linked to the previous and other nearby resources we
instead only keep track of when we enter and exit new resources'
geographic areas, using geographic boundary links, and then convert
those entry and exit times into larger reservations intervals that are
equivalent to merging together the multiple smaller intervals that the
naive algorithm produces.

It does require that we pre-calculate the boundary links between each
resource and its furthest away geographically linked resources but
such an operation is trivial to calculate when generating the
geographic links themselves. See Figure \ref{fig:subdivision_grid} for
an example of boundary resources when using adjacency geographic
linking.

Our proposed boundary algorithm becomes especially useful when
applying $s$ edge subdivisions as not only does the length of the
average time-path changes from $n$ to $sn$, since the increase in the
number of resources is linear in proportion to the number of edge
subdivisions, but the average number of geographically linked
resources for each resource also increases to $sm$. This leads to a
much more expensive $O(s^2nm)$ run-time complexity using
\cite{stenzel8}'s algorithm compared to our algorithm's $O(sn)$.

\begin{algorithm}
	\caption{Naive Algorithm}
	\label{alg:naive}
	\begin{algorithmic}
		\State \textbf{Given} time-path $T$
		\State \textbf{Given} geographic links between resources $L$
		\State \textbf{Initialise} output reservations $R \gets []$

		\For {every (resource, time-window) $(r, t) \in T$}
		\State append $(r, t)$ to $R$
			\For {every resource $p \in L[r]$}
				\State append $(p, t)$ to $R$
			\EndFor
		\EndFor
		\State \textbf{return} $R$
\end{algorithmic}
\end{algorithm}

\begin{algorithm}
	\caption{Boundary Algorithm}
	\label{alg:boundary}
	\begin{algorithmic}
		\State \textbf{Given} time-path $T$
		\State \textbf{Given} geographic links between resources $L$
		\State \textbf{Given} geographic boundary links between resources $B$
		\State \textbf{Initialise} output reservations $R \gets []$
		\State \textbf{Initialise} swallowed start times $S \gets \{\}$
		\State \textbf{Initialise} momentary reservations $V \gets \{\}$

		\Statex
		
		\State $(r, t) \gets$ the first (resource, time-window) $\in T$
		\For {every resource $p \in L[r]$}
			\State $V[p] \gets t$
		\EndFor

		\Statex

		\For {every (resource, time-window) $(r, [s1, e1]) \in$ the tail of $T$}
			\For {every (resource, time-window) $(p, [s2, e2]) \in V$}
				\If {$p \in L[r]$}
					\State $S[p] \gets s2$
				\Else
					\State append $(p, [s2, e2])$ to $R$
				\EndIf
			\EndFor

			\Statex

			\State clear $V$

			\Statex

			\For {every boundary resource $b \in B[r]$}
				\If {$b \in S$}
					\State $V[b] \gets [S[b], e1]$
					\State remove $b$ from $S$
				\Else
					\State $V[b] \gets [s1, e1]$
				\EndIf
			\EndFor
		\EndFor

		\Statex

		\State $end \gets$ end of the last time-window $\in T$
		\For {every (resource, time) $(p, s) \in S$}
			\State append $(p, [s, end])$ to $R$
		\EndFor
		\For {every (resource, time-window) $(p, t) \in V$}
			\State append $(p, t)$ to $R$
		\EndFor
		\State \textbf{return} $R$
\end{algorithmic}
\end{algorithm}

\section{Simulations}
\label{sec:simulations}

\subsection{Scheduling Decisions}
\label{sec:scheduling_decisions}

Any valid anchored time-pathing algorithm must make three main decisions in
order to complete a set of given demands. AGV-demand assignment
\cite{egbelu}, anchor node selection (related to idle AGV positioning
rules \cite{improta}), and the ordering in which to time-path AGVs
with their assigned demands \cite{mors11,rybecky18,rybecky20}.

The effects of different AGV-demand assignment, anchor selection, and
time-path ordering rules is not the study of this paper, hence, in our
simulations we use random AGV-demand assignment with random anchor
selection and random time-pathing ordering.

A secondary decision relevant to the online version of the AGV routing
problem is how to change the timetable upon the receipt of new
demands. The predominantly used approach is the conservative myopic
strategy \cite{qiu} in which all previously made assignments and
time-paths are respected and considered immutable. Since the effect of
different timetable update strategies is also not the study of this
paper, we employ the conservative myopic strategy for our simulations.

\subsection{Implementation Details}
\label{sec:implementation_details}

A library was built in the Rust programming language to perform the
described algorithms and simulations.

All simulations were performed on a computer equipped with an Intel
i5-12500.

\subsection{Graph Generation}
\label{sec:graph_generation}

\begin{figure}[h!]
\centering
\includegraphics[width=6cm]{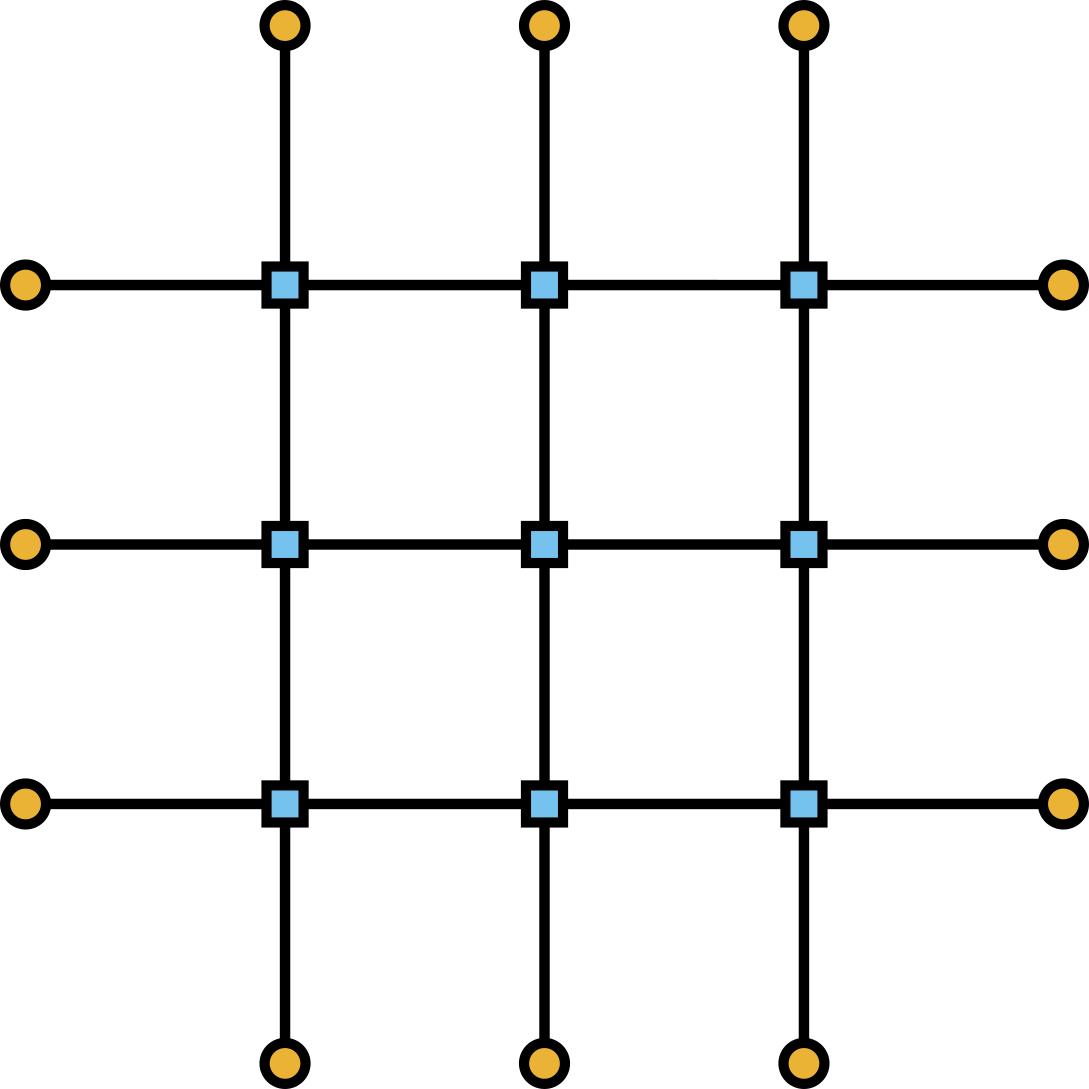}
\captionsetup{justification=centering}
\caption{A $4{\times}4$ grid graph. The orange circles are anchor
nodes whereas the blue squares are non-anchor nodes that can be used
for demands.}
\label{fig:4grid}
\end{figure}

Two-dimensional grid graphs were used for the simulations due to their
ease of implementation and their ability to be procedurally generated
at any size. The anchor nodes were set as nodes on the perimeter of
the grid, the corner nodes were removed, and finally, the edges
between anchor nodes were removed in order to satisfy the key
assumptions in section \ref{sec:modelling_and_key_assumptions}. An
example of a $4{\times}4$ grid graph is show in Figure \ref{fig:4grid}.

The weights of all edges in the graph were set to a time-unit of 5000,
this represents the distance of the edges divided by the speed of the
AGVs or how long it would take an AGV to travel the length of an edge.

\subsection{Anchorisation}
\label{sec:anchorisation_simulations}

To test the two anchorisation algorithms, greedy and naive, we
generated random placements of AGVs on a $100{\times}100$ grid graph and
measured their run-times. When selecting the random positions of AGVs
we also had to ensure no two AGVs were positioned on the same resource
as that would cause anchorisation to fail.

\begin{figure}[h!]
\centering
\includegraphics[width=8cm]{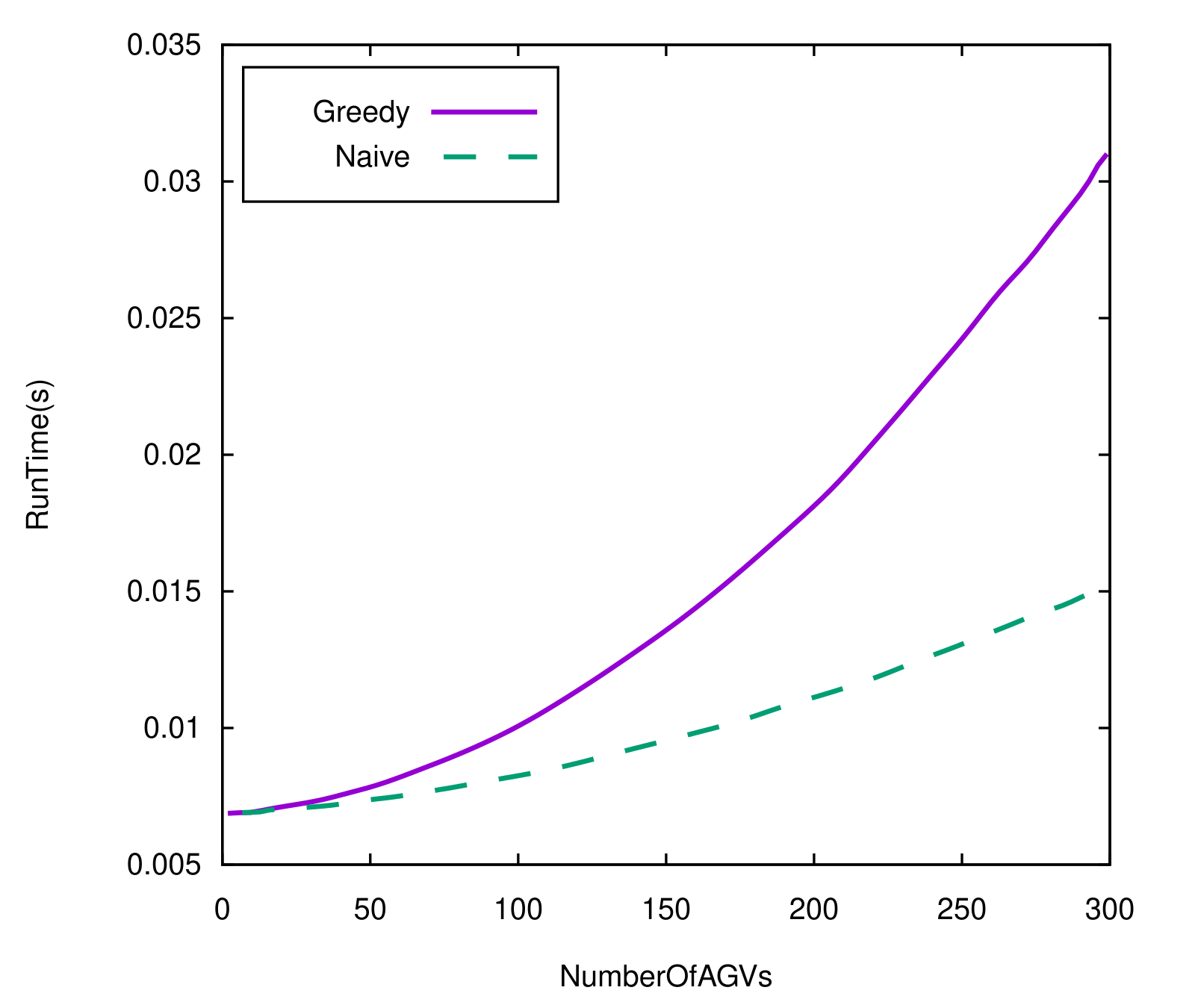}
\captionsetup{justification=centering}
\caption{Anchorisation algorithms against run-time for various numbers
of AGVs on a grid graph of size $100{\times}100$.}
\label{fig:anchorisers}
\end{figure}

Interestingly, the results shown in Figure \ref{fig:anchorisers} seem
in contrast with their respective worst-case run-time analyses since
the greedy algorithm is reasonably slower than the naive algorithm for
all numbers of AGVs. We believe the reason for this discrepancy is due
to the fact that the run-time analysis of the naive algorithm assumes
the worst-case scenario of each inner-loop search for an anchorisable
AGV to fail on every AGV except the last which would succeed. However,
in practise on the given grid graph we observed that this was rarely
the case and on average only one or two AGVs needed to be checked as
nearly every AGV had a unblocked physical path to an anchor node. On
the other hand the greedy algorithm would on average do much more
labelling per found time-path due to starting with labels at every
unanchored AGVs' positions and hence it took longer to find an
anchoring time-path for each AGV.

The discussed success rate of the naive algorithm is likely due to the
regularity of the grid graph layout which means very few bottleneck
resources that may cause failure if blocked by a static AGV. Hence the
performance of the two algorithms should be empirically benchmarked on
a case-to-case basis as the most performant algorithm is likely to
vary on different graph layouts and possibly many other factors.

\subsection{Time-Pathing Optimisations}
\label{sec:time_pathing_optimisations}

Perhaps the most important metric of this entire framework of
algorithms is the overall time taken to calculate a timetable for a
given number of AGVs with a given number of demands and on given
graph, since if it takes too long it will make it harder to implement
real-time systems capable of reacting to changing demands.

We simulated four different time-pathing algorithms: FullZero,
FullManhattan, PartialDijkstras and PartialManhattan. FullZero is our
implementation of the time-pathing algorithm with all the features
described in section \ref{sec:additional_time_pathing_features} but
with no guided search heuristic and no partial search since the entire
base graph is used. FullManhattan is the same as FullZero but uses a
manhattan distance heuristic guidance function. PartialDijkstras is
the same as FullZero but uses a chain-topology partial graph generated
by sequential Dijkstras searches on the base graph. Finally, the
PartialManhattan is the same as PartialDijkstras but uses manhattan
distance guided A* search instead of Dijkstras for generating the
partial graphs.

It is important to note that the manhattan distance heuristic is only
an admissible heuristic function in this situation since our graph is
grid-based and would have to be replaced with a real pre-calculated
graph-distance lookup table for equivalent performance with non-grid
graphs.

\begin{figure}[h!]
\centering
\includegraphics[width=8cm]{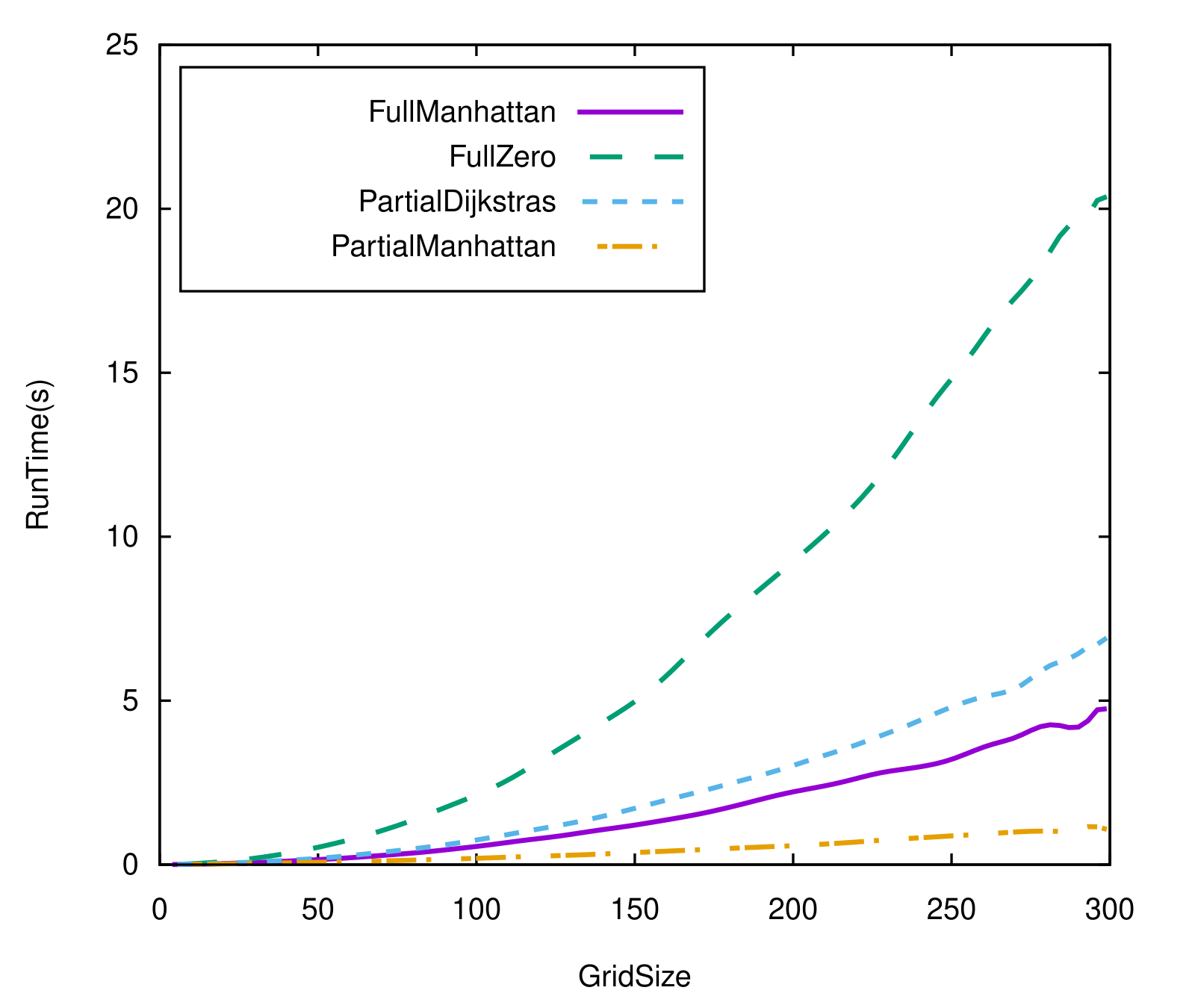}
\captionsetup{justification=centering}
\caption{Time-pathing algorithms against run-time for various
$n{\times}n$ grid graph sizes with 4 AGVs and 40 random demands.}
\label{fig:presets-runtime}
\end{figure}

As is shown in Figure \ref{fig:presets-runtime}, FullZero performs
significantly worse than the other algorithms due to it exploring and
labelling in every direction from the starting location until its
destination is found similar to Dijkstras. FullManhattan limits this
effect by prioritising labels spatially closer to the destination
which may not always be the best direction due to the temporal
dimension but it is a good predictor more often than not. The speedup
from FullZero to FullManhattan is somewhat similar to the traditional
speedup gained from Dijkstras to the A* algorithm in regular graph
searching. On the other hand, the two partial searching algorithms,
PartialDijkstras and PartialDijkstras, also perform significantly
better than FullZero with the only difference between them being the
graph search algorithm used for generating the partial graph. As such,
it can be concluded that the actual time spent time-pathing is closer
to the time taken by PartialManhattan.

However, it is a valid concern that the use of partial search although
beneficial for run-time may have a detrimental effect on solution
quality. Therefore, we measured two basic metrics to quantify solution
quality, total makespan (the total time taken to complete all demands
and for all AGVs to anchor) and the total distance travelled by all
AGVs. These two metrics approximate two major considerations when
evaluating any material handling system, total throughput (which is
roughly the inversely proportional to makespan) and total running
costs (assuming that running costs are a function of total AGV travel
distance).

\begin{figure}[h!]
\centering
\includegraphics[width=8cm]{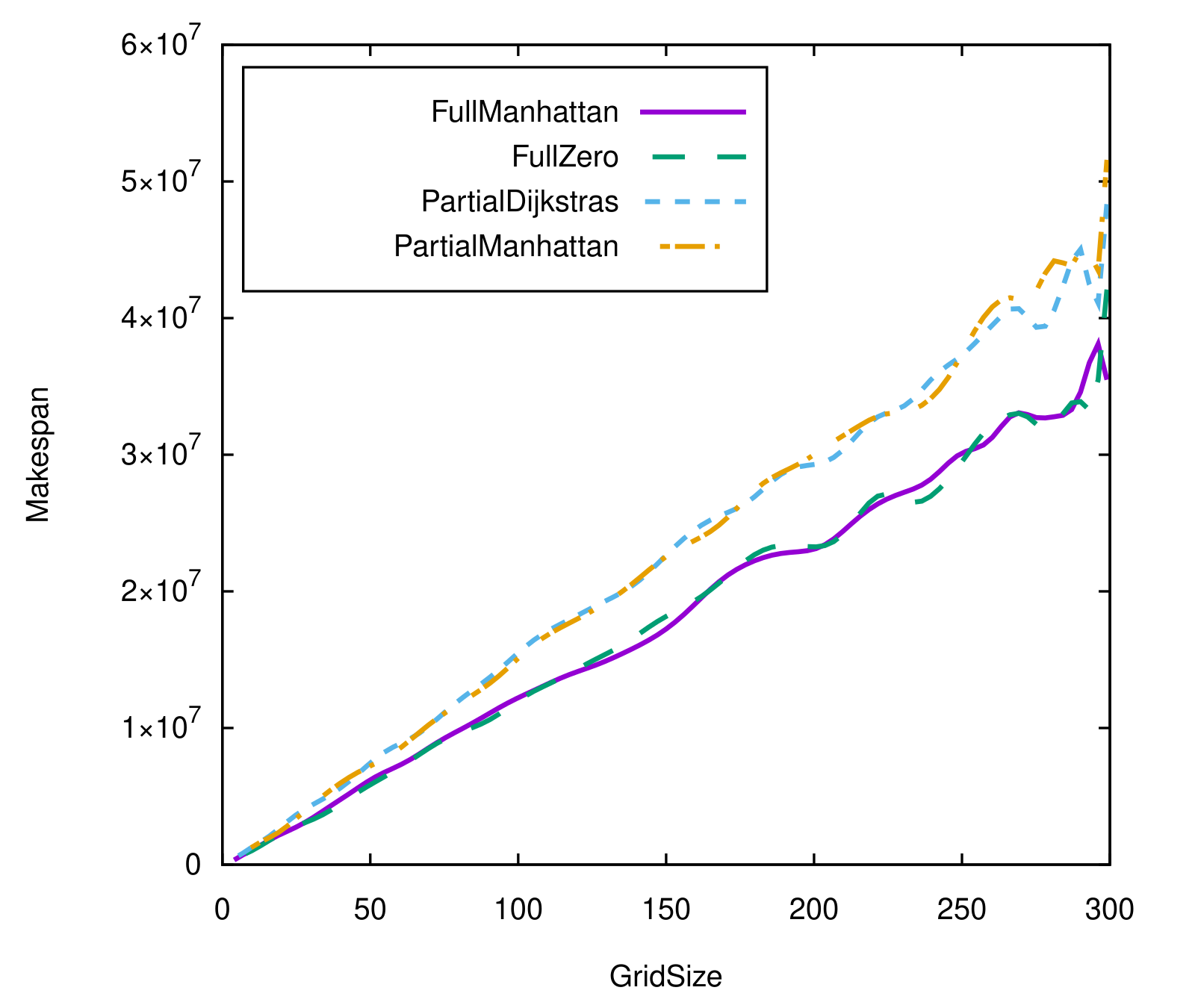}
\captionsetup{justification=centering}
\caption{Time-pathing algorithms against the total makespan for
various $n{\times}n$ grid graph sizes with 4 AGVs and 40 random
demands.}
\label{fig:makespan}
\end{figure}

As can be seen in Figure \ref{fig:makespan}, there is a slight drop in
makespan between the Full to Partial algorithms. The amount of
difference in makespan between Full and Partial Time-Pathing
algorithms is quite small here with respect to the total makespan
however this proportion and the difference between them is also likely
to depend on the graph layout, the number of AGVs and other scheduling
decisions so empirical testing on a case by case basis is advisable.

\begin{figure}[h!]
\centering
\includegraphics[width=8cm]{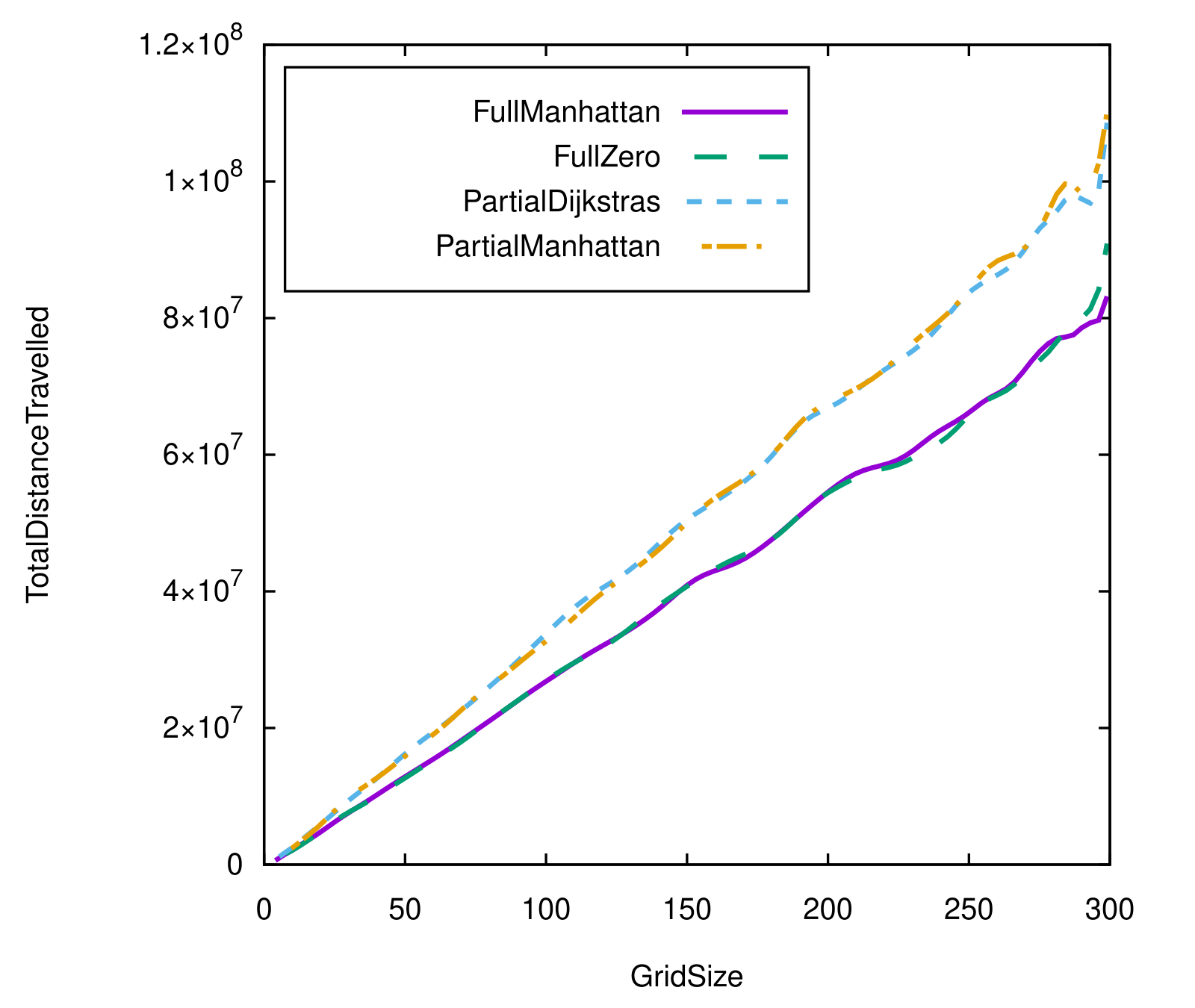}
\captionsetup{justification=centering}
\caption{Time-pathing algorithms against the total disance travelled
by all AGVs for various $n{\times}n$ grid graph sizes with 4 AGVs and
40 random demands.}
\label{fig:presets-total-distance-travelled}
\end{figure}

Figure \ref{fig:presets-total-distance-travelled} shows a very similar
scenario to the makespan between the Full and Partial algorithms. This
is likely the case since the grid graph used is fairly regular and so
due to the lack of any significant bottlenecks in the graph most of
the time spent by an AGV, when assigned a demand, is actively
travelling not waiting and so the makespan and total distance
travelled are directly proportional. This would likely not always be
the case on more congested graphs however.

\subsection{Geographic Reservation Algorithms}
\label{sec:geographic_reservation_algorithms}

\begin{figure}[h!]
\centering
\includegraphics[width=6cm]{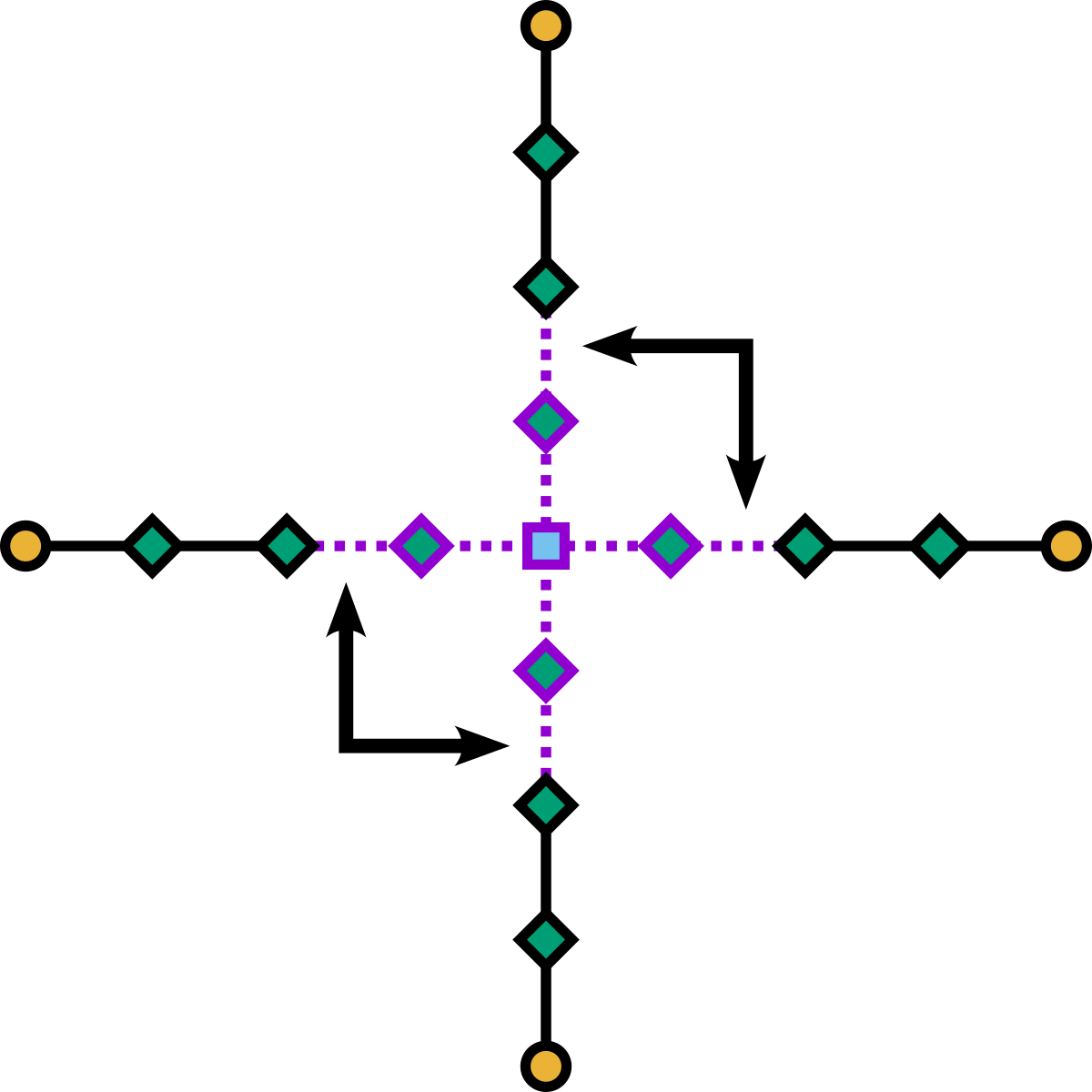}
\captionsetup{justification=centering}
\caption{A $2{\times}2$ grid graph that has been subdivided three
	times. The orange circles are anchor nodes, the blue square is a
	non-anchor node and the green diamonds are non-anchor nodes added
	during edge subdivision. The dotted purple edges and nodes are the
resources geographically linked to the centre node when using an
adjacency geographic linking of size three. The arrows point toward
the boundary resources which happen to all be edges in this example.}
\label{fig:subdivision_grid}
\end{figure}

Finally, we showcase a simulation of \cite{stenzel8}'s geographic
reservation algorithm vs our proposed boundary algorithm by measuring
the run-time of the two algorithms for a single time-path from one
corner to the opposite corner of a $100{\times}100$ grid graph with
varying amounts of graph subdivision. We use adjacency geographic
linking that scales linearly with the number of edge subdivisions so
the average distance of graph geographically linked to each resource
will roughly remain the same. For example, if we subdivided each edge
of a $2{\times}2$ grid graph three times then we would also
geographically link each resource to all resources less than or equal
to three adjacent resources away as is shown in Figure
\ref{fig:subdivision_grid}. In practise it may be more useful to use
AGV dimension polygons when working with two-dimensional graphs to
construct the geographic links as show in \cite{stenzel4,stenzel8} to
ensure no collisions are possible.

\begin{figure}[h!]
\centering
\includegraphics[width=8cm]{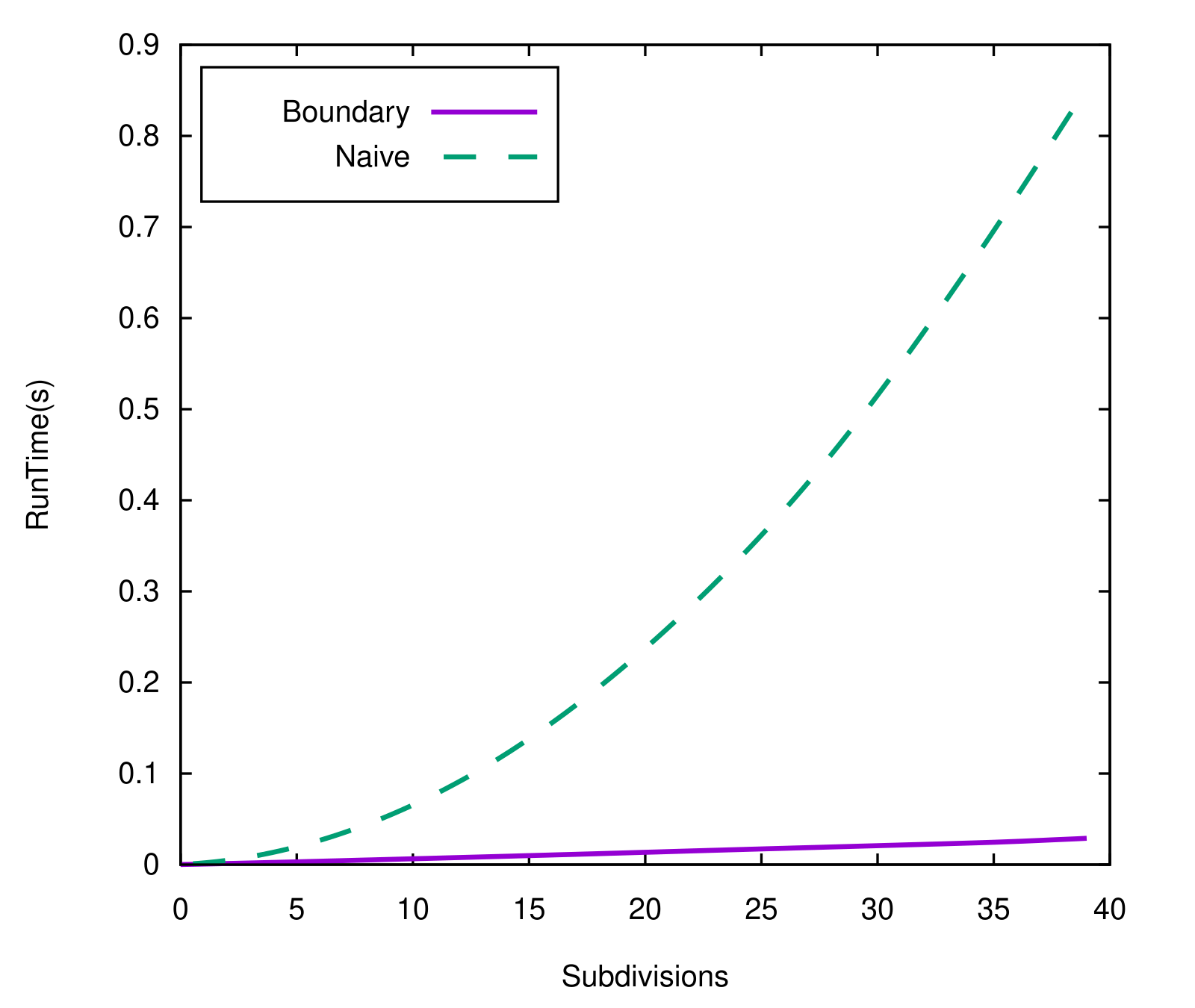}
\captionsetup{justification=centering}
\caption{Boundary vs Naive Geographic reservation algorithms against
run-time for various amounts of subdivision on a time-path from one
corner of a $100{\times}100$ grid graph to the opposite corner.}
\label{fig:reservers}
\end{figure}

As can be seen in Figure \ref{fig:reservers}, there are obvious
run-time complexity improvements from $O(s^2nm)$ to $O(sn)$ and since
there are no failure-probabilities involved, unlike with the
anchorisation run-time complexities, this results in a massive
run-time speedup. This speedup can have a large effect over all the
time-pathing needed to generate a single timetable, especially for
heavily geographically linked graphs or when using large amounts of
edge subdivision.

\printbibliography

\end{document}